\documentclass{PoS}

\usepackage{amssymb}

\title{Nuclear PDFs Today}

\ShortTitle{Nuclear PDFs Today}

\author{Hannu Paukkunen\\
        University of Jyvaskyla, Department of Physics, P.O. Box 35, FI-40014 University of Jyvaskyla, Finland \\
        Helsinki Institute of Physics, P.O. Box 64, FI-00014 University of Helsinki, Finland \\
        E-mail: \email{hannu.paukkunen@jyu.fi}}

\abstract{I review the current status of global analysis of nuclear parton distributions and discuss some near-future prospects. I also advocate proton-nucleus (p-$A$) measurements at the LHC with $A\ll 208$ from the viewpoint of nuclear parton distributions.}

\FullConference{International Conference on Hard and Electromagnetic Probes of High-Energy Nuclear Collisions\\
		30 September - 5 October 2018\\
		Aix-Les-Bains, Savoie, France}

\begin{document}

\section{State of the art 2018}

\vspace{-0.30cm}
A proper description of hard processes in high-energy collisions involving nuclei --- whether they take place in laboratory circumstances or in nature --- requires knowledge of nuclear parton distributions (PDFs). Table~\ref{tab:nPDFs} summarizes the latest available global analyses for these objects. The situation has been quite static for almost two years now, the most recent parametrizations being nCTEQ15 \cite{Kovarik:2015cma} and EPPS16 \cite{Eskola:2016oht}. Datawise, EPPS16 is the most comprehensive including e.g. LHC Run-I p-Pb data on dijets and electroweak bosons, as well as data on neutrino-nucleus deeply-inelastic scattering. On the theory side, the analyses at an NNLO precision are emerging and the heavy-quark effects are being taken into account, i.e. the PDFs are defined in general-mass variable flavour number schemes (GM-VFNS). For a long time, the light-quark flavour dependence was essentially neglected in the nuclear-PDF analysis, but now we are also making progress there and beginning to fold out the nuclear PDFs truly flavour by flavour. 
\vspace{-0.30cm}
\begin{table}[htb!]
\caption{Key specifications of contemporary nuclear-PDF analyses}
\vspace{-0.20cm}
\label{tab:nPDFs}
\begin{center}
{\renewcommand{\arraystretch}{1.1}
\begin{tabular}{|c||c|c|c|c|c|c|}
\hline
& \textsc{eps09} \cite{Eskola:2009uj} & \textsc{dssz12}	\cite{deFlorian:2011fp} & \textsc{ka15}	\cite{Khanpour:2016pph}	& \textsc{ncteq15} \cite{Kovarik:2015cma} & \textsc{epps16} \cite{Eskola:2016oht} \\
\hline
 {DIS in $\ell^-$+A}               & \checkmark           & \checkmark          & \checkmark     & \checkmark  & \checkmark        \\
 {Drell-Yan in p+A}              & \checkmark           & \checkmark          & \checkmark     & \checkmark  & \checkmark        \\
 {RHIC pions d+Au}               & \checkmark           & \checkmark          &                &  \checkmark & \checkmark        \\
 {$\nu$-nucleus DIS}          &                      & \checkmark          &                &             & \checkmark        \\ 
 {Drell-Yan in $\pi$+$A$}        &                      &                     &                &             & \checkmark        \\                                            
 {LHC p+Pb dijets        }       &                      &                     &                &             & \checkmark        \\
 {LHC p+Pb W, Z        }         &                      &                     &                &             & \checkmark        \\
 \hline
 \hline
 {Order in $\alpha_s$}           & NLO &  NLO           &  NNLO               &  NLO & NLO        \\
{$Q$ cut in DIS}                & $1.3 \, {\rm GeV}$   & $1 \, {\rm GeV}$    & $1 \, {\rm GeV}$ & $2 \, {\rm GeV}$ & $1.3 \, {\rm GeV}$   \\
 {datapoints}                    & 929                  & 1579                & 1479             & 708       & 1811               \\
 {free parameters}               & 15                   & 25                          & 16                        & 16   & 20               \\
 {error tolerance }       & 50                      & 30                          & N.N                        & 35      & 52            \\
 {proton baseline}          & {\textsc{cteq6.1}}   &  {\textsc{mstw2008}}    &  {\textsc{jr09}} &  {\textsc{cteq6m}-like} & {\textsc{ct14nlo}}\\
 {GM-VFNS}           &                    & \checkmark                     &                    & \checkmark  & \checkmark           \\
 {flavour separation}            &                       &                         &                       & valence & valence + sea \\
\hline
\end{tabular}
}
\end{center}
\end{table}

\begin{figure}[htb!]
\includegraphics[width=0.33\linewidth]{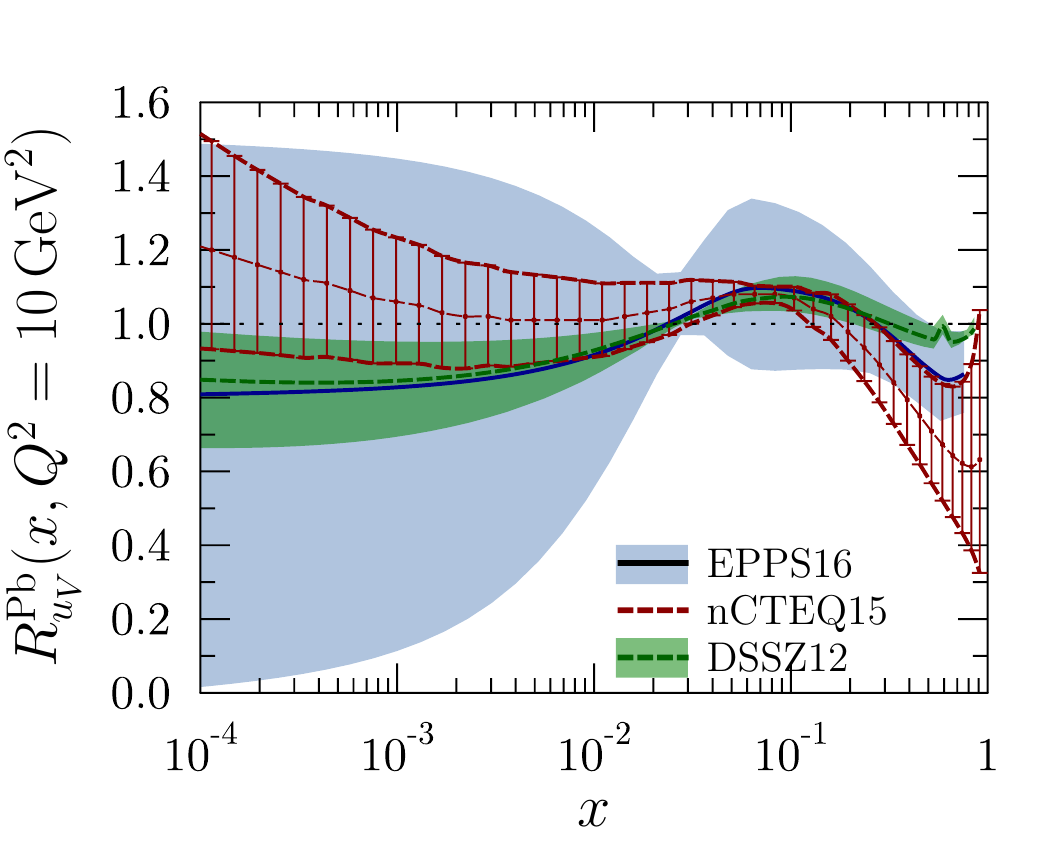}
\includegraphics[width=0.33\linewidth]{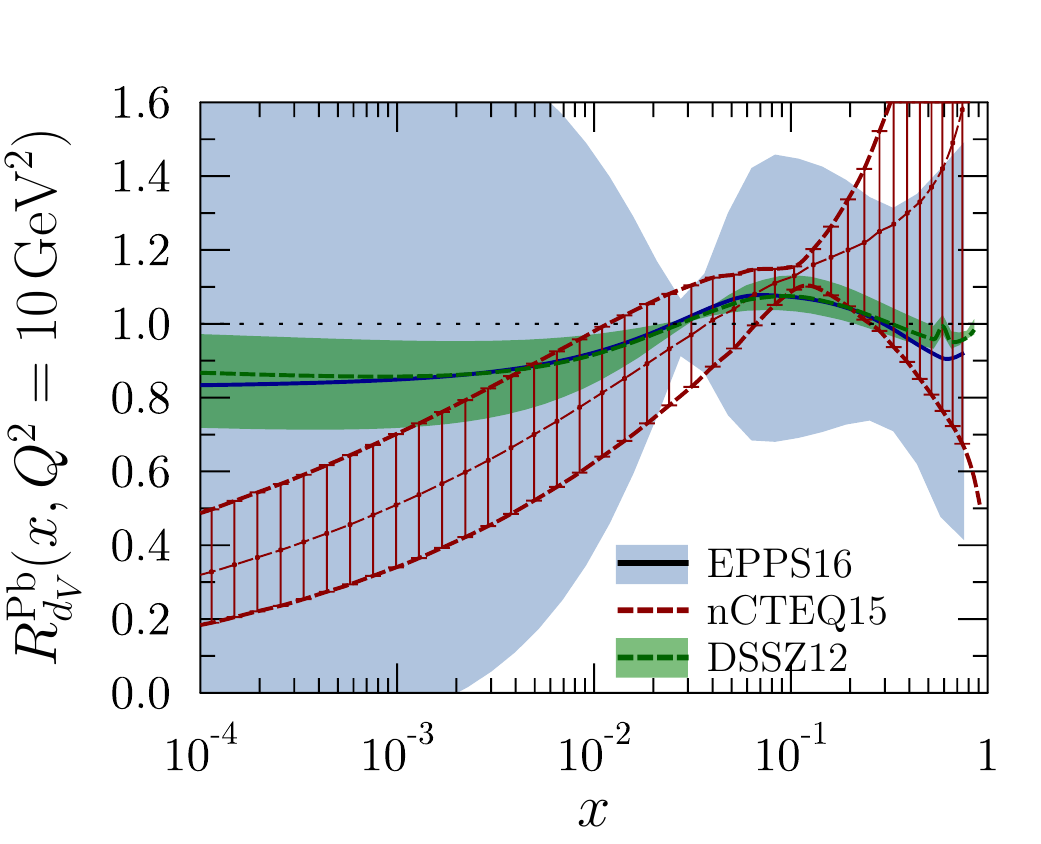}
\includegraphics[width=0.33\linewidth]{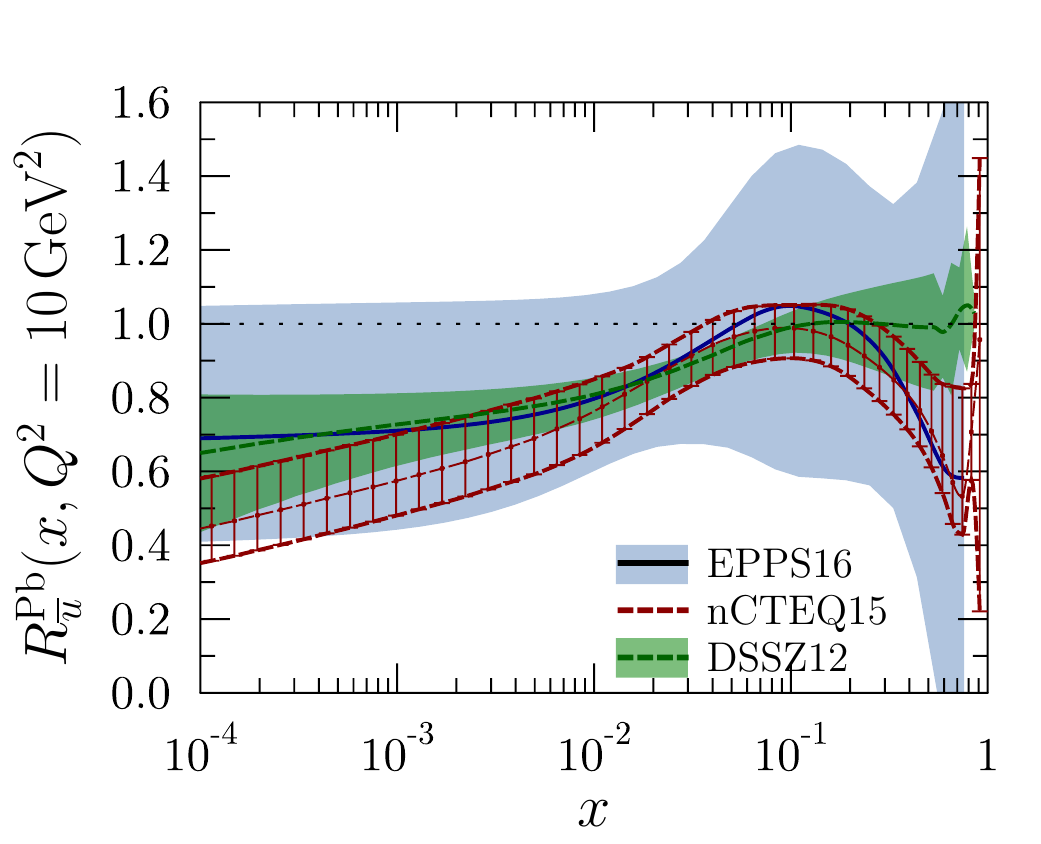} \\
\includegraphics[width=0.33\linewidth]{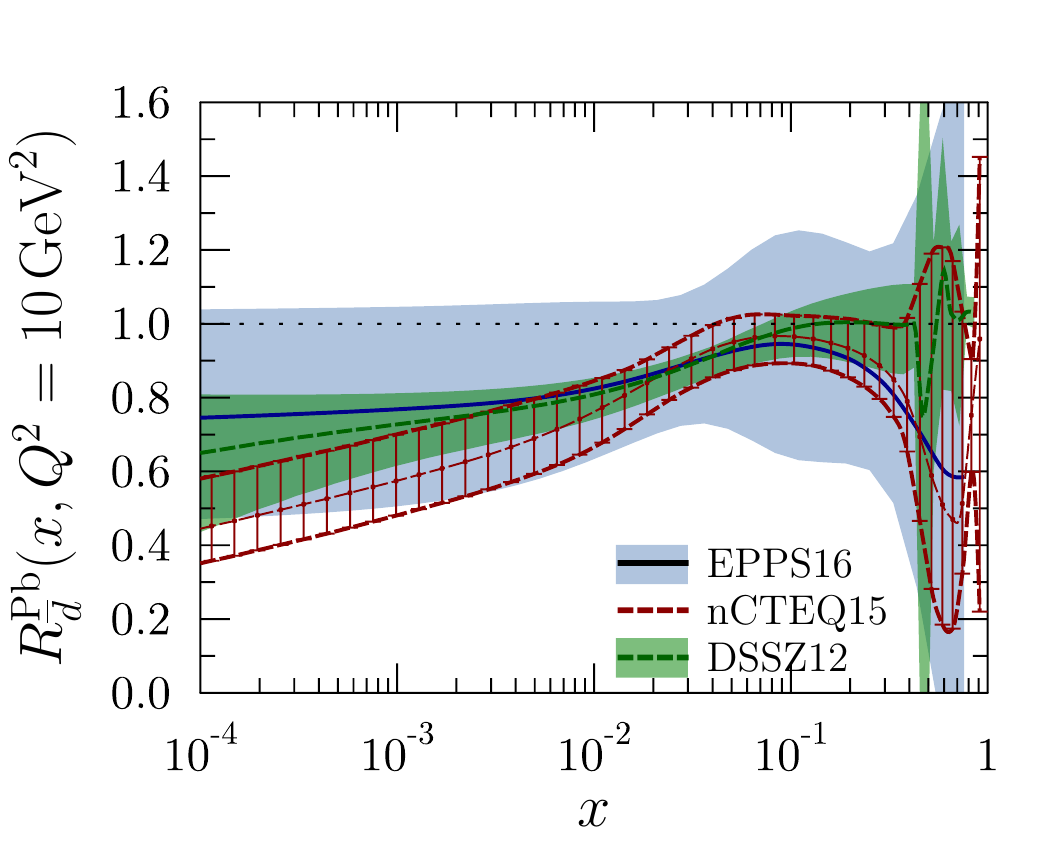} 
\includegraphics[width=0.33\linewidth]{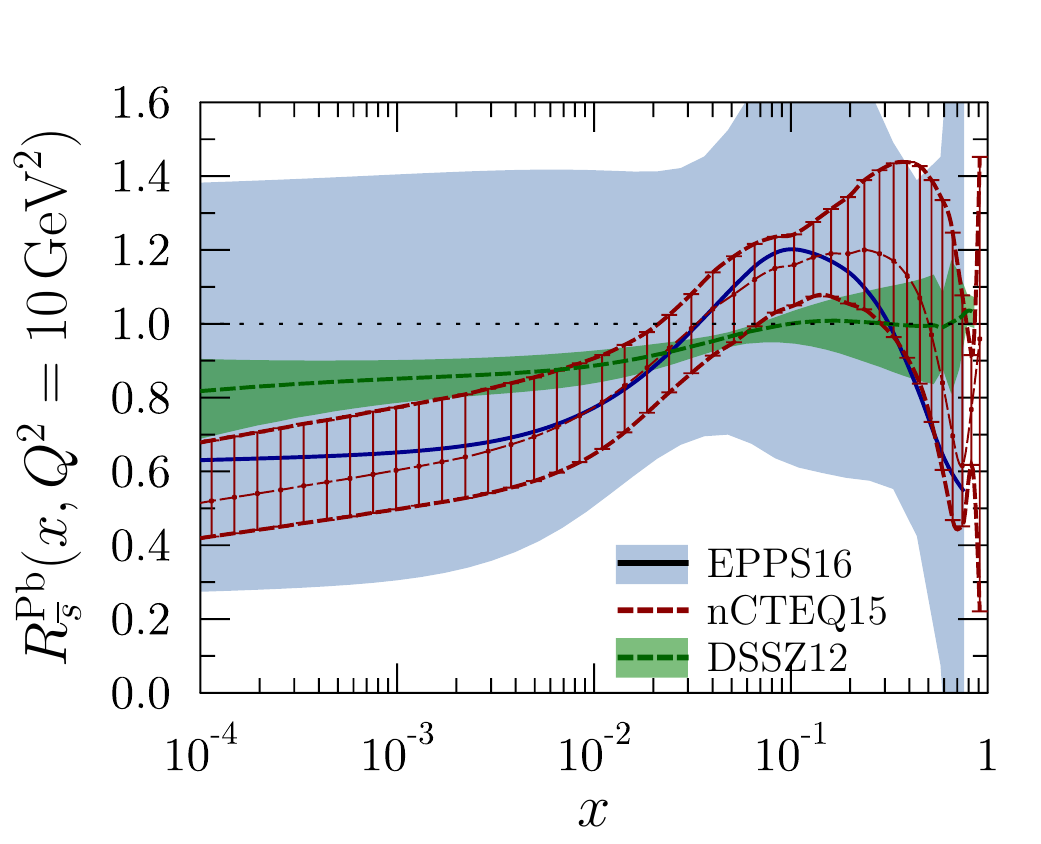} 
\includegraphics[width=0.33\linewidth]{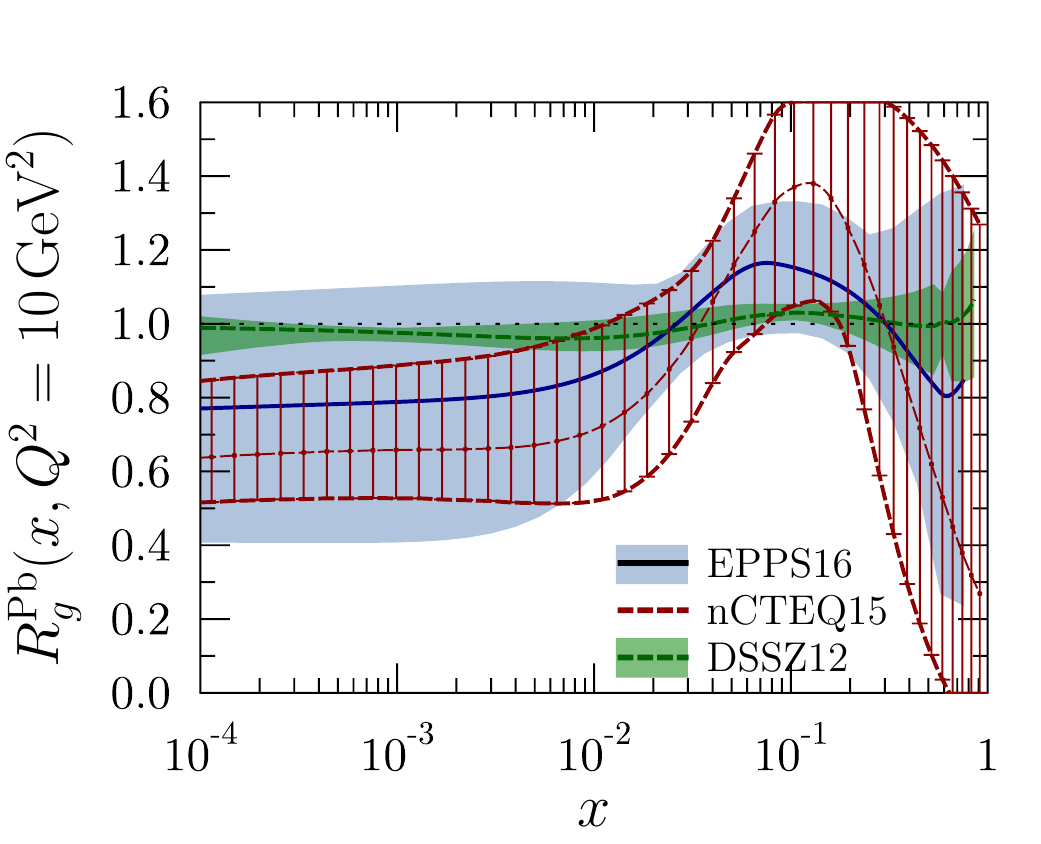} \vspace{-0.6cm}
\caption{Nuclear modifications from the EPPS16, nCTEQ15 and DSSZ global NLO fits at $Q^2=10\,{\rm GeV}^2$.}
\label{fig:AllnPDFs}
\end{figure}

\begin{figure}[htb!]
\includegraphics[width=0.33\linewidth]{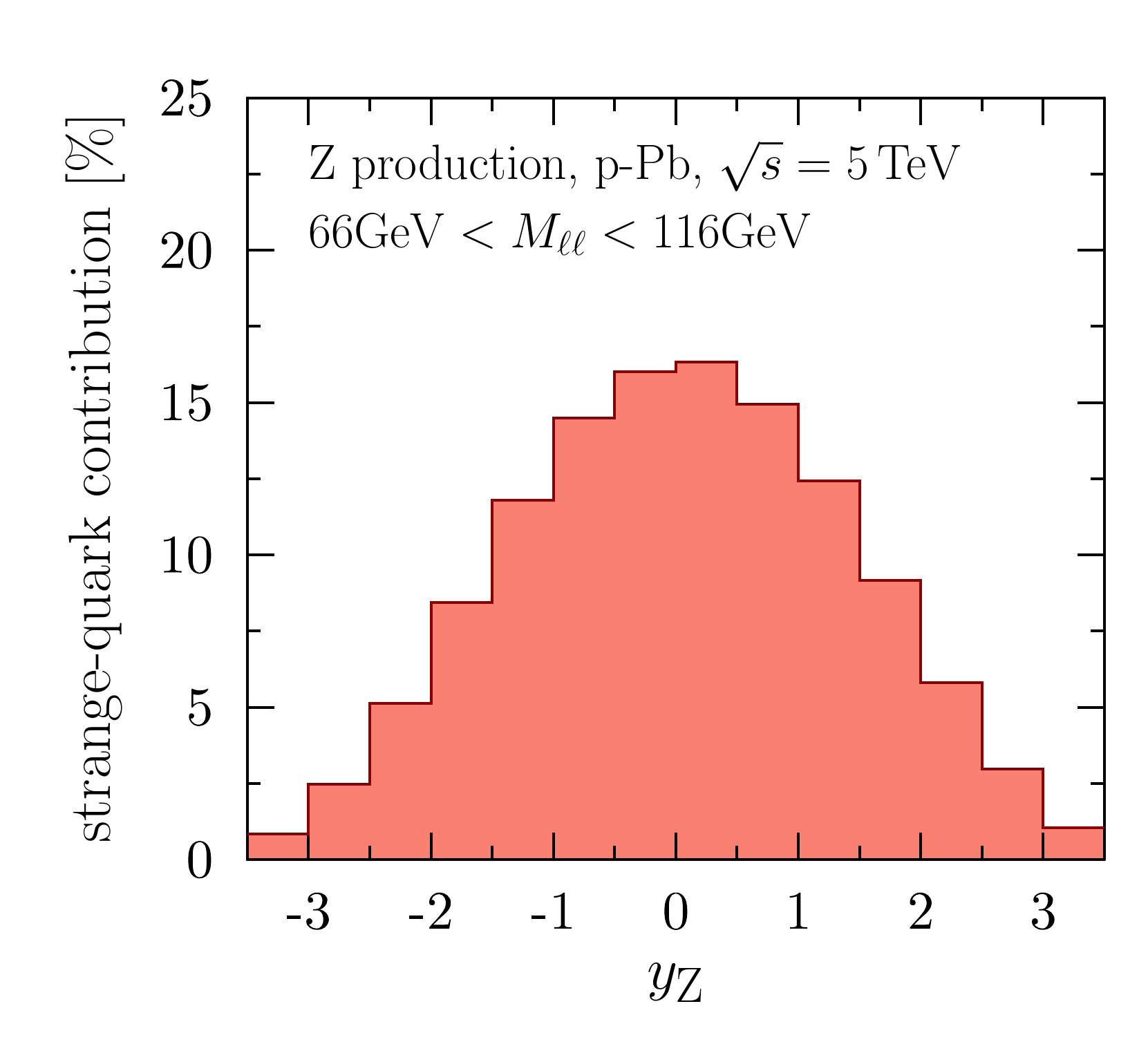}

\vspace{-4.65cm}\hspace{4.99cm}
\includegraphics[width=0.33\linewidth]{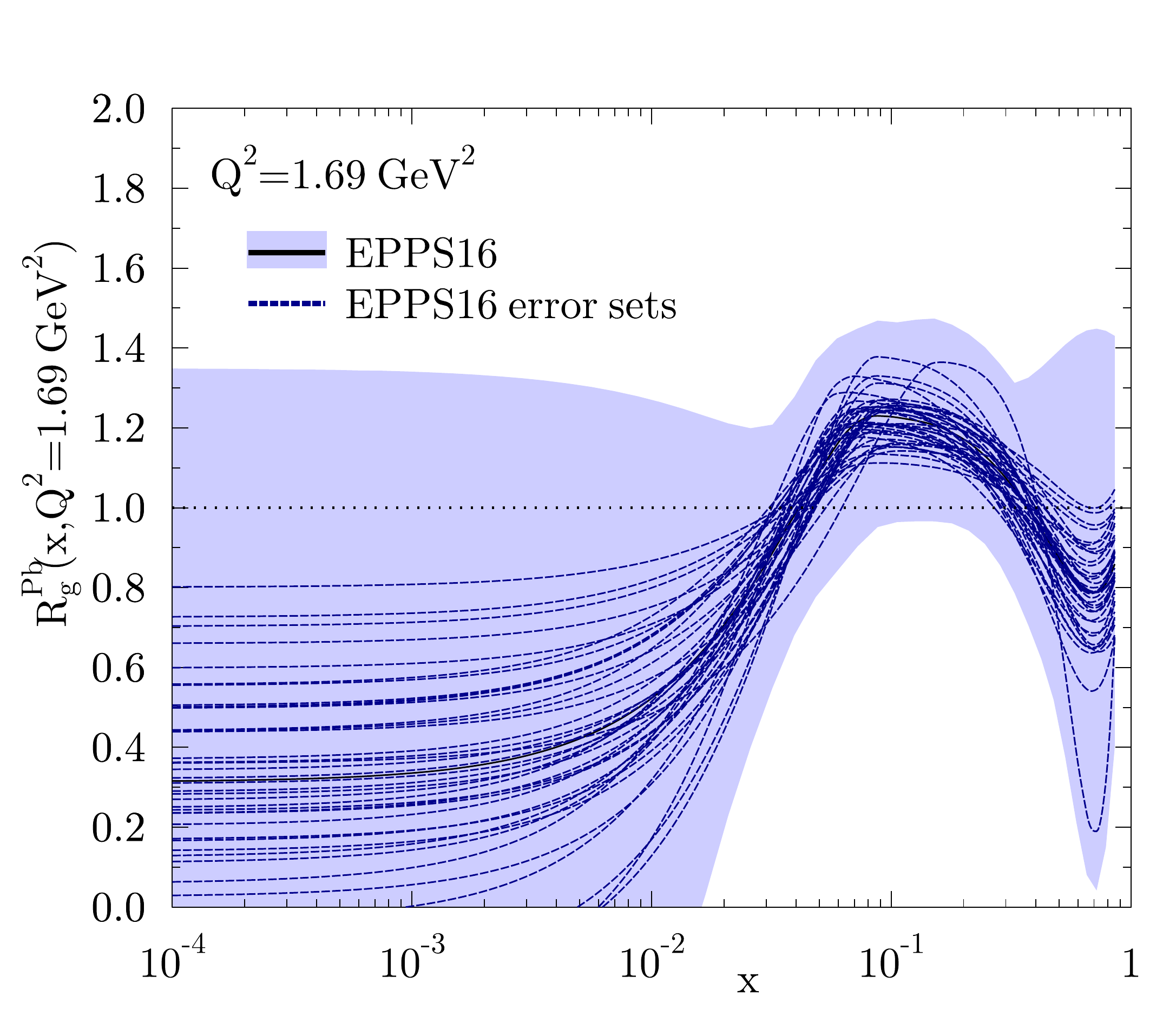}
\includegraphics[width=0.33\linewidth]{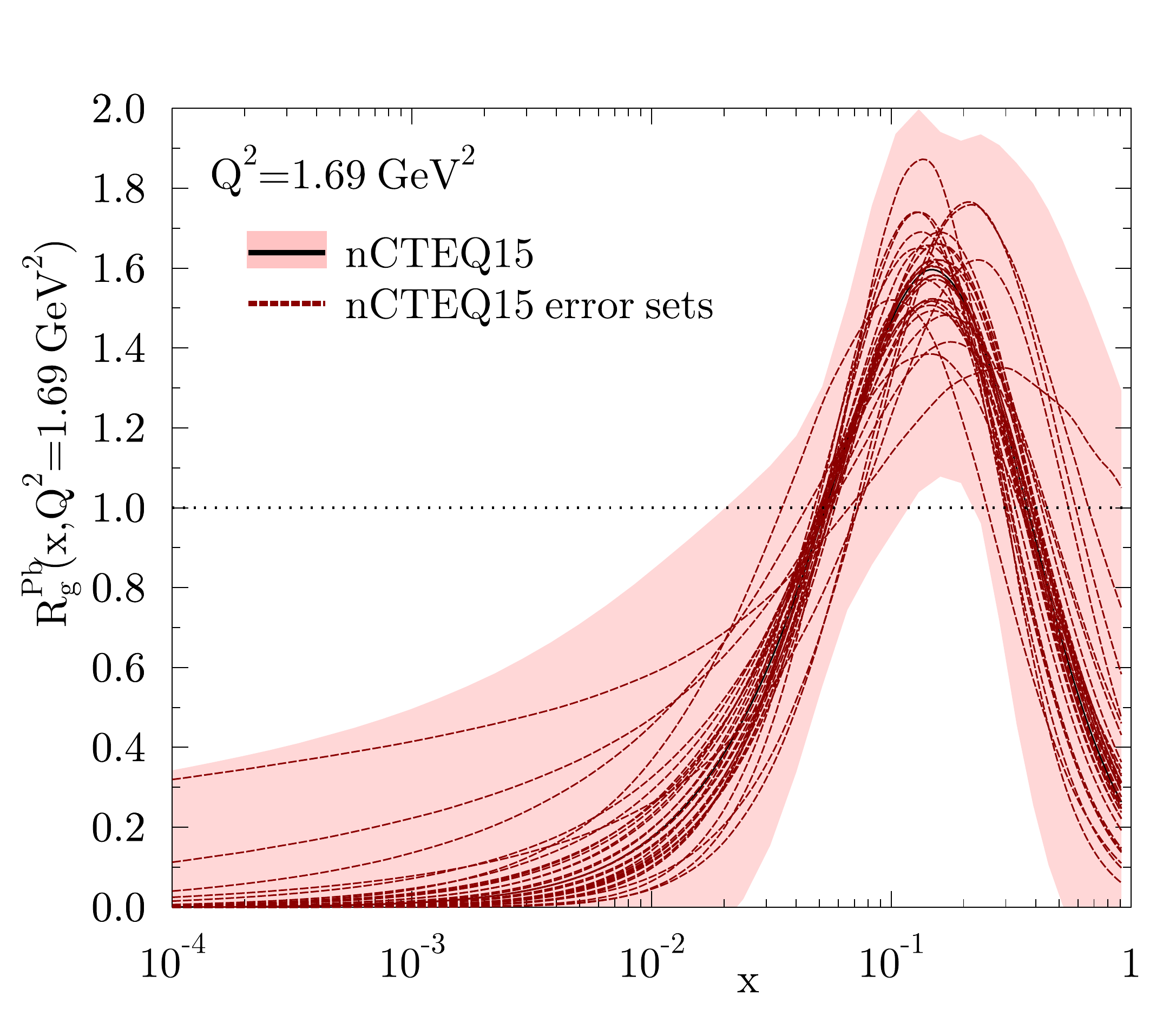}
\caption{{\bf Left:} Percentual contribution of $s\overline{s}$ scattering to Z production at $\sqrt{s}=5\,{\rm TeV}$.
{\bf Middle:} The EPPS16 gluon nuclear modifications with individual error sets at $Q^2=1.69\,{\rm GeV}^2$.
{\bf Right:}  The nCTEQ15 gluon nuclear modifications with individual error sets at $Q^2=1.69\,{\rm GeV}^2$.
}
\label{fig:s}
\end{figure}

\vspace{-0.6cm}
Let us have a look on the most recent NLO extractions. To this end, Figure~\ref{fig:AllnPDFs} presents a comparison of the PDF nuclear modifications $R_i^A(x,Q^2)$ defined as ratios between the free- and bound-proton PDFs,
\vspace{-0.1cm}
\begin{equation}
 R_i^A(x,Q^2) \equiv f_i^{{\rm proton \ in \ nucleus \ } A}(x,Q^2) / f_i^{{\rm free \ proton}}(x,Q^2) \,,
\end{equation}
from EPPS16, nCTEQ15 and DSSZ12. {\bf Valence Quarks:} In the EPPS16 and nCTEQ15 analyses the up- and down-valence quarks were independently parametrized. Whereas in EPPS16 the central values for u and d come out mutually similar --- thanks e.g. to the neutrino-Pb DIS data --- in nCTEQ15 the up valence has a very strong high-x EMC effect, while the down-valence is enhanced at large $x$. Within the uncertainty bands, however, EPPS16 and nCTE15 are still consistent. In the DSSZ12 analysis there was no flavour freedom, and thus the uncertainties are artificially much smaller. {\bf Sea Quarks:} In the EPPS16 analysis the three light sea-quark flavours (u,d,s) were independently parametrized while in nCTEQ15 and DSSZ there is no such freedom. Thus the EPPS16 uncertainties are larger, but on the other hand there is less bias. The largest uncertainties are there for the strange quarks, for which the constraints are rather scarce. Currently the best constraints for the strange come from the Z-boson production at the LHC where there is around 20\% contribution from $s\overline{s}$ scattering at midrapidity, as illustrated in the leftmost panel in Figure~\ref{fig:s}. {\bf Gluons:} At large $x$, the nCTEQ15 uncertainty bands are wider than those of EPPS16. This is principally due to the LHC dijet data included in the EPPS16 analysis, leading to a better constrained large-$x$ gluon. At low $x$, the nCTEQ15 uncertainties are smaller, which is probably just related to the form of the fit functions --- I will elaborate more on this below. In DSSZ12 there are practically no nuclear effects in gluons, as the authors assumed nuclear modifications in parton-to-pion fragmentation functions when analyzing the RHIC pion data. 

\vspace{-0.4cm}

\vspace{-0.1cm}
Concerning the functional forms used to fit nuclear PDFs. What we need are certain assumptions for the $x$ and $A$ dependence for $R_i^A(x,Q^2)$ at the parametrization scale $Q^2_0$ --- the behaviour at higher $Q^2$ follows then from the evolution equations. The assumed functional forms are rather simple and --- as has been shown in Refs.~\cite{Aschenauer:2017oxs,Paukkunen:2017phq} --- the bias in the current parametrizations is huge, particularly at small $x$. As can be understood from the right-hand panels of Figure~\ref{fig:s}, not much variation in small-$x$ behaviour of $R_i^A(x,Q^2)$ is allowed by the current parametrizations. The difficulty is not so much in inventing a flexible ansatz for the $x$ dependence, but to do it in such a way that also the $A$ dependence can be made physically sound (``larger effects for a larger nucleus''). In this sense nuclear PDFs are more difficult to fit than the free proton PDFs. There is also an interesting ongoing effort to fit the nuclear PDFs in a neural-network framework, which should be superior when it comes to reducing the parametrization bias \cite{Khalek:2018bbv}. 

\vspace{-0.65cm}

\section{New constraints from the LHC}

\begin{figure}[htb!]
\centering
\includegraphics[width=0.95\linewidth]{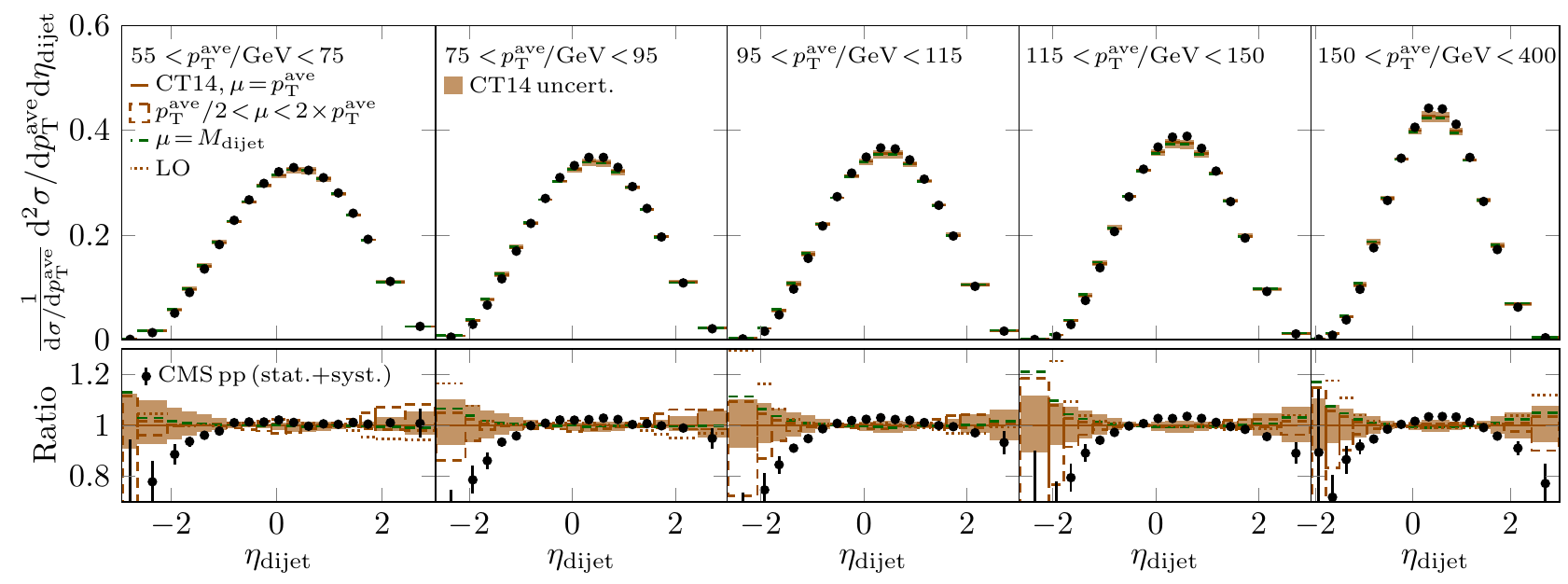}
\includegraphics[width=0.95\linewidth]{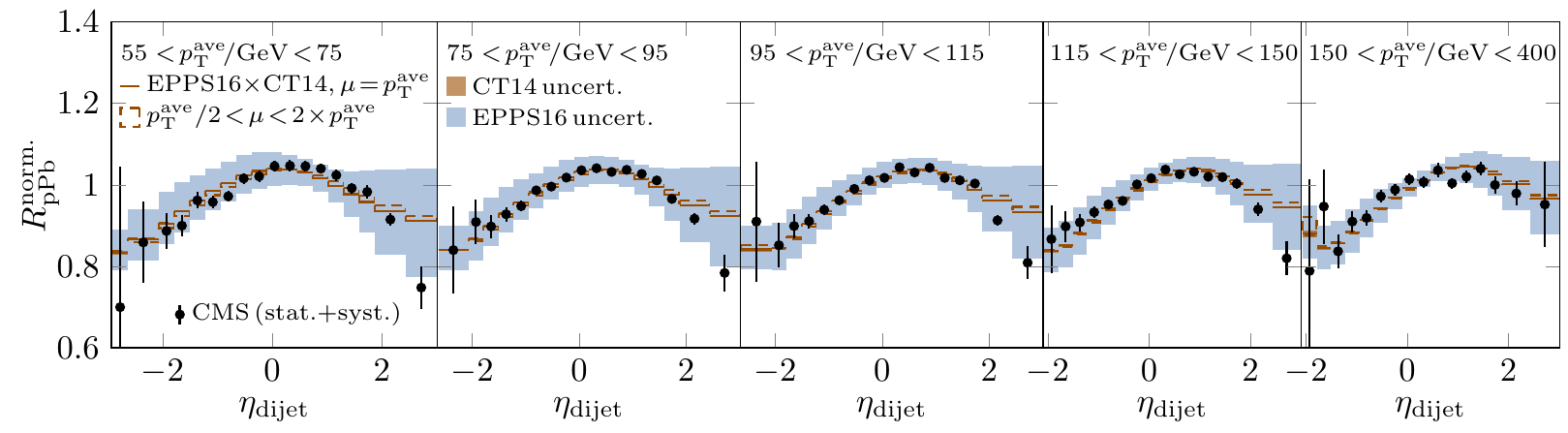}
\caption{{\bf Up:} The CMS dijet spectrum in p-p compared with the NLO calculations using CT14NLO proton PDFs \cite{jetpaper}. {\bf Down:}
Measured dijet nuclear modification compared with EPPS16 predictions \cite{jetpaper}.}
\label{fig:CMSdijet}
\end{figure} 

\vspace{-0.1cm}
{\bf Dijets:} One of the most stringent new PDF constraints in the horizon are the CMS 5-TeV dijet data \cite{Sirunyan:2018qel}. The upper panels in Figure~\ref{fig:CMSdijet} show the p-p baseline data for the normalized yields
\vspace{-0.1cm}
$$
\frac{d\sigma^{\rm pp}(\eta_{\rm dijet},p_{\rm T}^{\rm ave})}{\int d\sigma^{\rm pp}(\eta_{\rm dijet},p_{\rm T}^{\rm ave})d\eta_{\rm dijet}} \,, \,\, \eta_{\rm dijet} \equiv \frac{\eta_{\rm leading}+\eta_{\rm subleading}}{2} \,, \,\, p_{\rm T}^{\rm ave} \equiv \frac{p_{\rm T}^{\rm leading}+p_{\rm T}^{\rm subleading}}{2} \,,
$$

\vspace{0.1cm}
\noindent where $\eta_i$ refers to the jet pseudorapidity. A remarkable thing is that the data cannot be well described within the uncertainties of the current NLO proton PDFs --- the theory predictions are way too broad. This observable, apart from the large $|\eta_{\rm dijet}|$ tails, should not be too sensitive to the NNLO corrections either, so there seems to be a problem. However, the problem appears to be the same in the p-Pb case, and upon taking the p-Pb/p-p ratio, the outcome is rather well described by EPPS16, as shown in the lower panels of Figure~\ref{fig:CMSdijet}. The large-$\eta_{\rm dijet}$ data appear to prefer a stronger shadowing that is possible within EPPS16. Nevertheless, the data uncertainties are way smaller than the EPPS16 error bands, and in this sense these data promise to have a major impact on the gluon PDFs. Preliminary dijet measurements by ATLAS \cite{ATLAS:2018imx} have also recently appeared.

{\bf Z, $\gamma$ and W$^\pm$:} On-shell electroweak bosons are clean probes of PDFs at high factorization scales \cite{Paukkunen:2010qg,Kusina:2016fxy,Ru:2016wfx}. However, the Run-I data included in the EPPS16 analysis still had a relatively small statistical weight as only a handful of data points was available. For the higher luminosity and higher $\sqrt{s}$, the new 8-TeV data are now significantly more precise and begin to discriminate between the available sets of nuclear PDFs \cite{CMS:2018ilq}. Also, the data uncertainties are of the same order as the NNLO QCD corrections, motivating the NNLO-level analyses of nuclear PDFs. Now that the yields are higher, it may also be possible to start realistically looking into e.g. the $p_{\rm T}$ distributions of the electroweak bosons which can provide complementary information with respect to the $p_{\rm T}$-integrated case \cite{Guzey:2012jp,Brandt:2014vva}. The direct photons are also good probes of the gluon density \cite{Helenius:2014qla}. As the cross sections for lowish $p_{\rm T}$ photons are much much larger than e.g. for Z and W$^\pm$, the statistics are high. The limiting factor seems to be currently the systematic experimental uncertainty \cite{ATLAS:2017ojy}.

\begin{figure}
\centering
\vspace{-0.8cm}\includegraphics[width=0.49\linewidth]{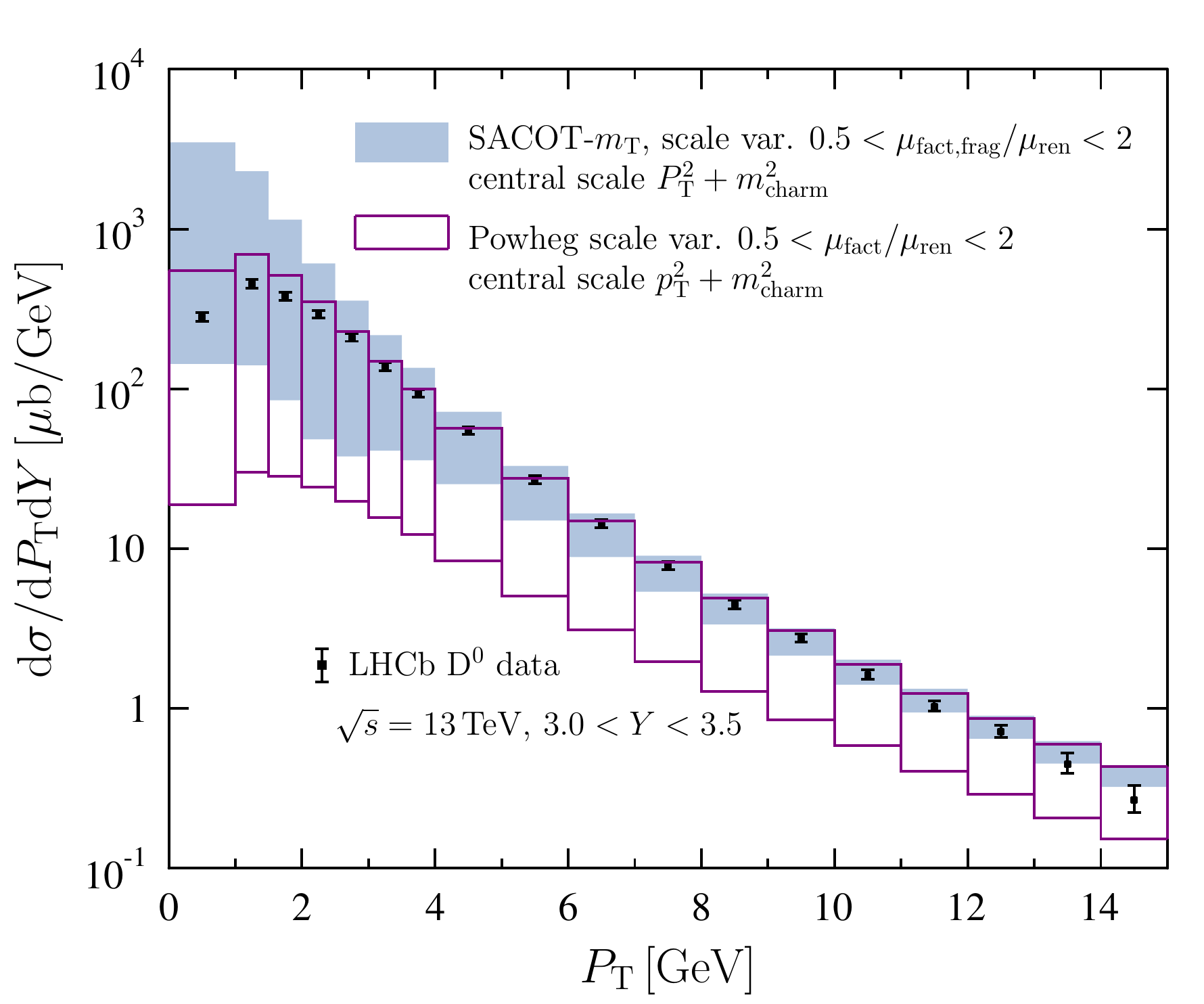}
\includegraphics[width=0.49\linewidth]{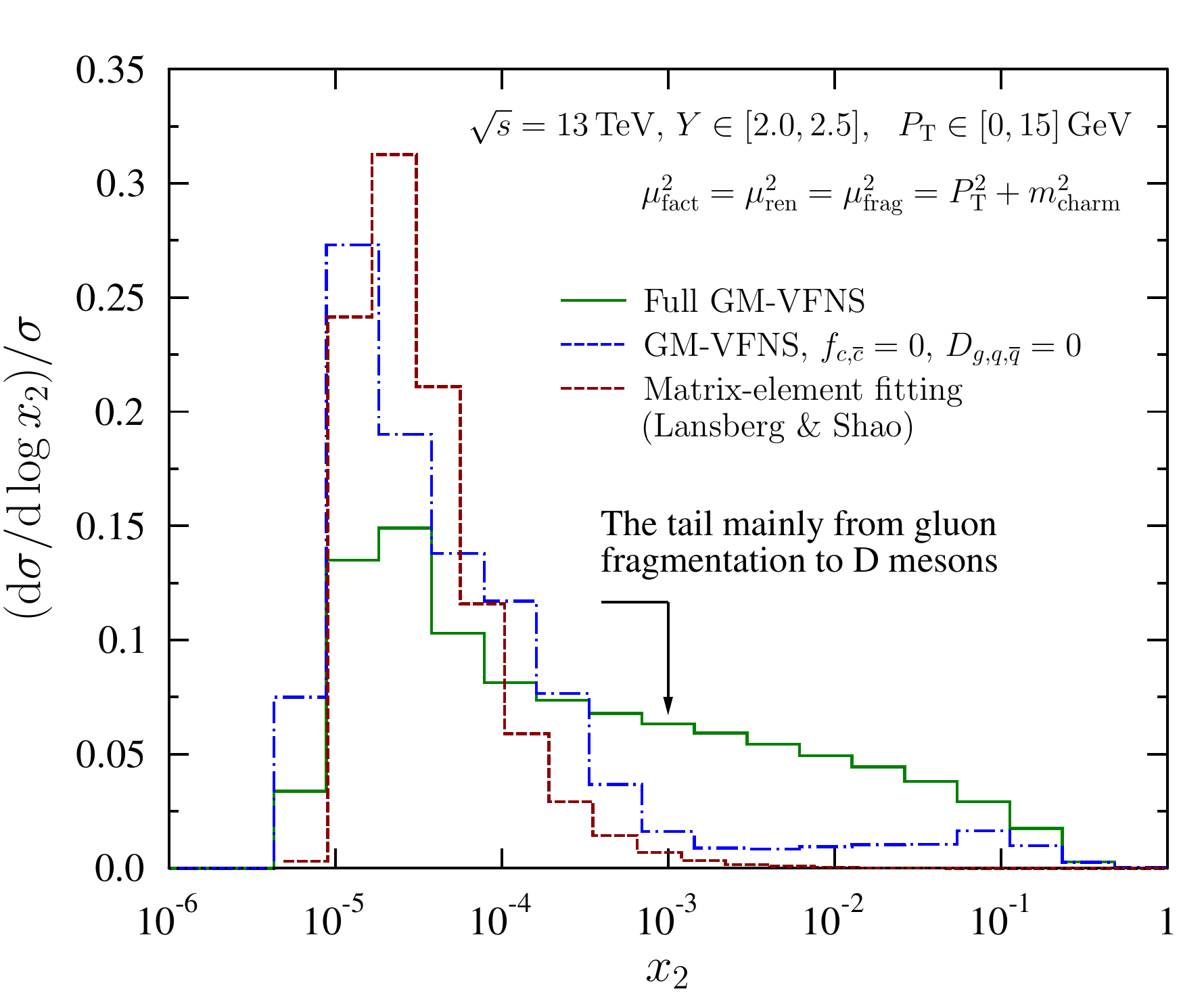}
\caption{{\bf Left:} Comparison of the LHCb 13-TeV D$^0$ data \cite{Aaij:2015bpa} with theoretical GM-VFNS and Powheg-method calculations. The error bands represent the scale uncertainties. Figure is from Ref.~\cite{Helenius:2018uul}. {\bf Right:} The $x_2$ distributions corresponding to LHCb kinematics from GM-VFNS and matrix-element-fitting approaches.}
\label{fig:Dspectrum}
\end{figure}

{\bf Open heavy flavour:} The open heavy flavours, D mesons in particular, have recently attracted a growing interest, and we have already rather precise data from various LHC experiments \cite{Acharya:2017hdv,ATLAS:2017dgr,Aaij:2017gcy}. In particular, the LHCb data \cite{Aaij:2017gcy} at forward direction show a compelling evidence of shadowing at clearly perturbative scales. From the PDF viewpoint the challenge is, maybe a bit surprisingly, in the theoretical treatment for which there are several options, including the fixed-order perturbative QCD \cite{Mangano:1991jk}, FONLL \cite{Cacciari:1998it}, and GM-VFNS frameworks \cite{Kniehl:2004fy,Helenius:2018uul}. As most of the general-purpose nuclear PDFs are defined within GM-VFNS, where the heavy quarks treated as partons above the mass thresholds, this framework would sound like the most adequate one. And it is not only that. It is rather well known that the other frameworks, e.g. FONLL and even sophisticated fixed-order calculations matched to parton showers (``\textsc{powheg}+\textsc{pythia}'') \cite{Frixione:2007nw}, tend to undershoot the LHC (and other) data by a factor of two or so, though within the large scale uncertainties an agreement can be reached. The GM-VFNS framework, on the other hand, reproduces the data very well and at high $p_{\rm T}$ has a much smaller scale uncertainty than e.g. \textsc{powheg}+\textsc{pythia} setup. An example is shown in the left-hand panel of Figure~\ref{fig:Dspectrum}.

\begin{figure}[htb!]
\centering
\vspace{-0.2cm}
\includegraphics[width=0.485\linewidth]{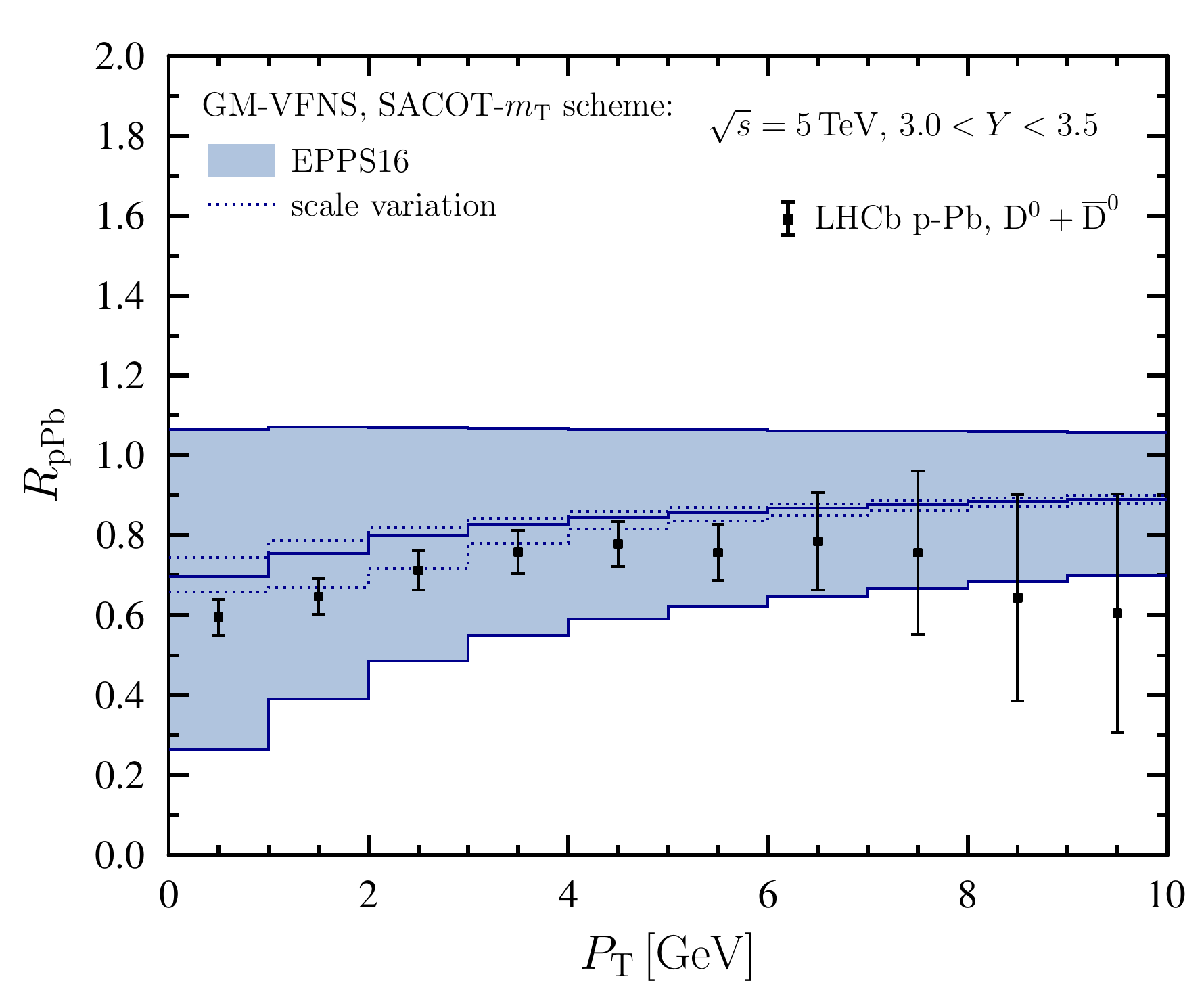}
\includegraphics[width=0.485\linewidth]{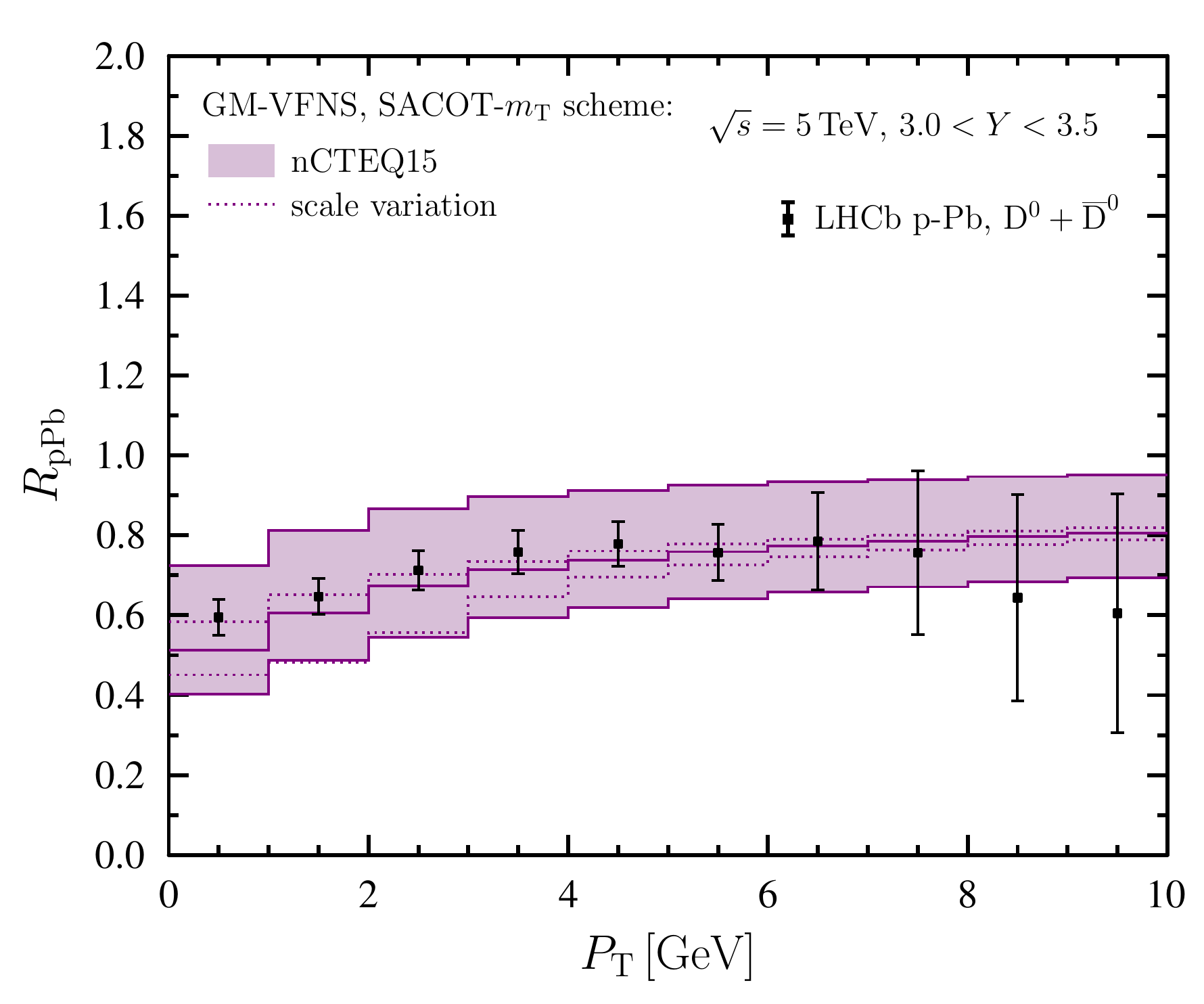}
\caption{LHCb measurements on the nuclear modification $R_{{\rm pPb}}$ for D$^0$ mesons compared with GM-VFNS calculations using EPPS16 (left) and nCTEQ15 (right) \cite{DPaper}.}
\vspace{-0.3cm}
\label{fig:DmesonRpA}
\end{figure} 

A very different strategy was adopted in a recent article by Lansberg and Shao \cite{Lansberg:2016deg}, where the idea was not to calculate the heavy-flavour cross sections from the first principles but rather to write the cross sections as
$
d\sigma^{{\rm D}^0} = f_g(x_1,Q^2_f) \otimes d\hat\sigma_{gg}^{{\rm D}^0} \otimes f_g(x_2,Q^2_f),
$            
parametrize $d\hat\sigma_{gg}^{{\rm D}^0}$, and fit it to available p-p data. The p-Pb predictions are then obtained by simply replacing the proton PDF by a nuclear PDF. I call this here ``matrix-element fitting''. As was shown in a follow-up paper \cite{Kusina:2017gkz}, one can obtain a very good description of e.g. the D-meson data by this construction. It works actually astonishingly well considering that the $x$ dependence of the assumed squared matrix elements was rather trivial, $\overline{|\mathcal{M}_{gg\rightarrow D^0 + X}|^2} = {\rm constant} \times x_1 x_2$, and that only the gluon-gluon scattering with leading-order kinematics was considered. Here, I should also point out that it was not just D mesons that were studied, but using the same framework it was possible to describe e.g. the inclusive J/$\psi$ production without invoking any energy loss, multiple scattering or equivalent. However, returning to the D-meson case, given the simple assumed form for the partonic cross sections, one may wonder whether it is really able to well enough capture the main features of the full-fledged NLO calculations. In the right-hand panel of Figure~\ref{fig:Dspectrum}, we see typical $x_2$ distributions corresponding to D mesons at forward LHCb kinematics from NLO GM-VFNS, and matrix-element-fitting method. The two distributions are quite different: GM-VFNS predicts a long tail towards large $x$ whereas the ansatz made in the matrix-element fitting yields a distribution which is more small-$x$ concentrated. Actually, the long tail in GM-VFNS comes mainly from the gluon fragmentation contributions \cite{Helenius:2018uul}, as suggested by the curves in Figure~\ref{fig:Dspectrum} with charm-quark PDFs and gluon/light-quark fragmentation contributions turned off. Thus, even for $R_{{\rm p}A}$, in which many theory uncertainties cancel, it is not indifferent how the cross section are computed. Of course, the big picture is qualitatively similar in these two frameworks, and also GM-VFNS predicts a significant shadowing at forward direction and the data are consistent with both, EPPS16 and nCTEQ15 as shown in Figure~\ref{fig:DmesonRpA}.

{\bf Pions, $\bf h^+h^-$:} There are also LHC p-Pb data on production of other species like neutral pions \cite{Acharya:2018hzf} or just unidentified charged hadrons \cite{Khachatryan:2015xaa}. All these are also sensitive to nuclear PDFs. For the presence of baryons in any $h^+h^-$ sample, the charged-hadrons can be used only at sufficiently high $p_{\rm T}$, $p_{\rm T} \gtrsim 10\,{\rm GeV}$ near midrapidity \cite{dEnterria:2013sgr} (could be lower at forward/backward directions), and there e.g. the data on foward-backward asymmetry are consistent with NLO predictions \cite{Khachatryan:2015xaa}. In the case of pions one should be able to go to lower values of $p_{\rm T}$. There, e.g. the ALICE data \cite{Acharya:2018hzf} agree with the EPPS16 predictions but disagree with nCTEQ15, thus promising some discrimination power. 

{\bf Top quarks:} The top-quarks have also finally been observed in p-Pb collisions, by CMS \cite{Sirunyan:2017xku}, and the theory-vs-data pattern for the total cross section looks similar as in p-p. The current statistics are, however, not enough to set significant constraints on nuclear PDFs \cite{dEnterria:2015mgr}. Thus, for the moment, the top production remains more a ``control'' measurement rather than a precision constraint.

\begin{figure}
\includegraphics[width=0.45\linewidth]{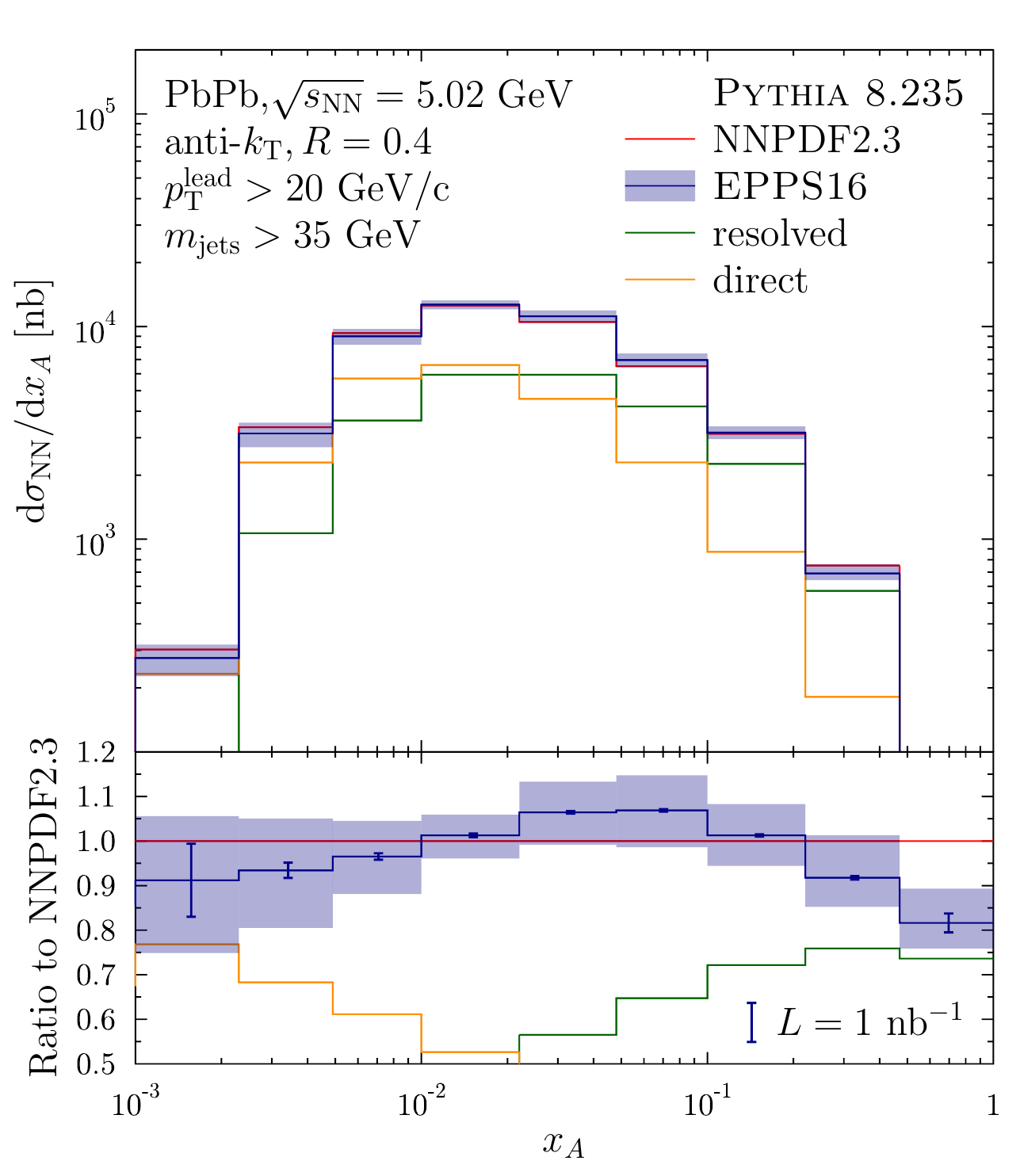}

\vspace{-8.70cm} \hspace{7.0cm}
\includegraphics[width=0.40\linewidth]{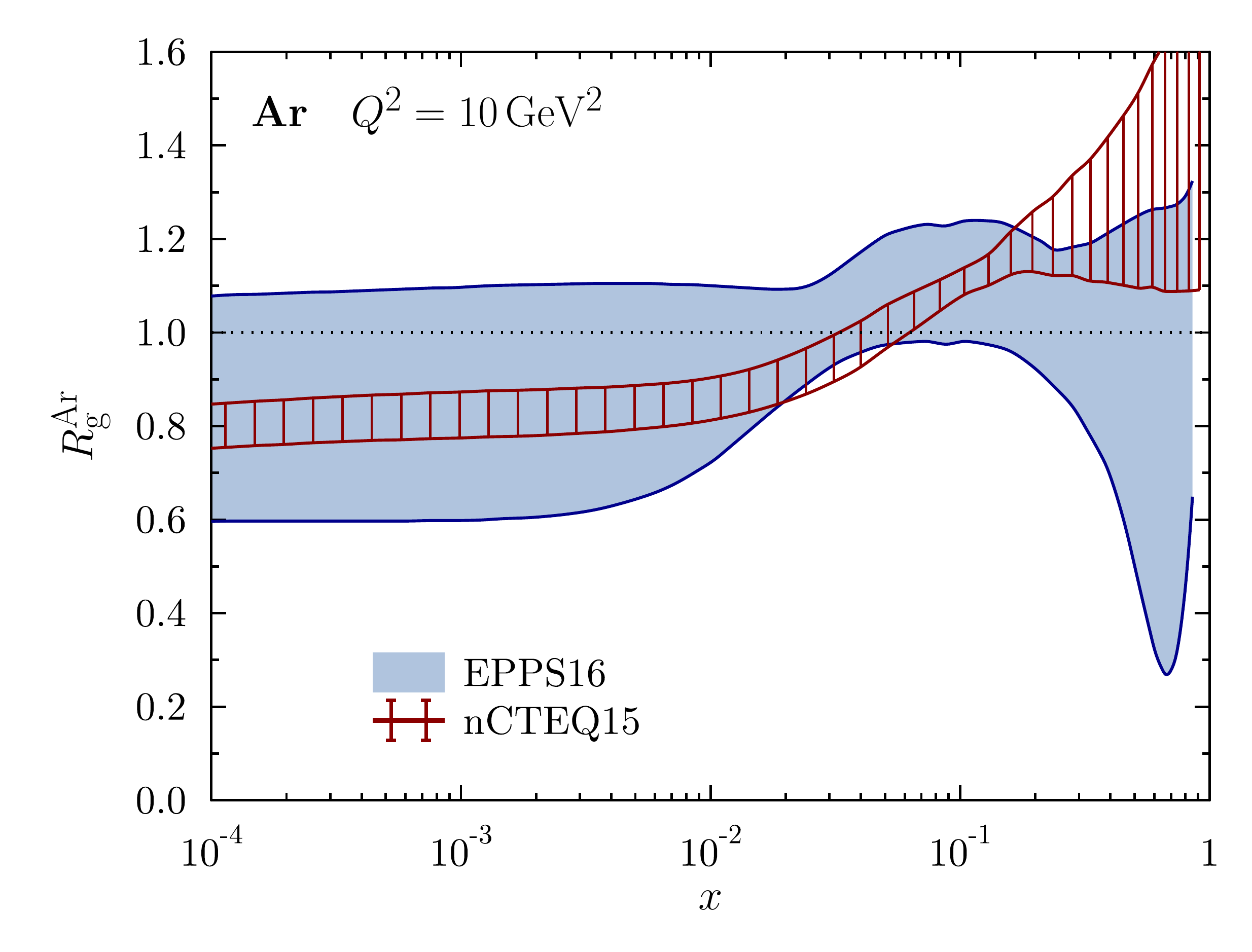}

\vspace{-0.10cm} \hspace{7.0cm}
\includegraphics[width=0.40\linewidth]{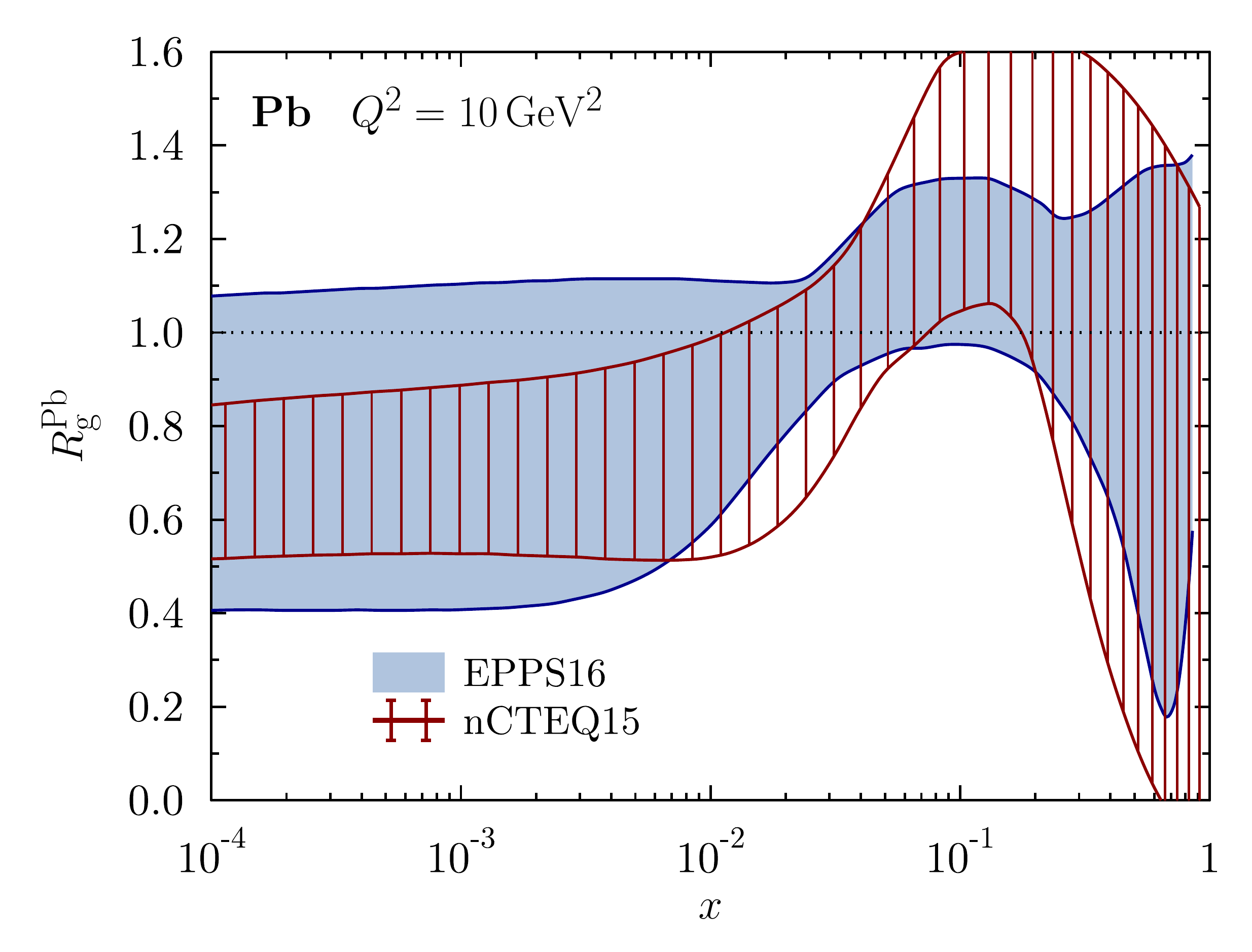}
\vspace{-0.30cm}
\caption{{\bf Left:} A \textsc{pythia} simulation of dijet photoproduction in ultraperiheral Pb-Pb collisions (from Ref.~\cite{Helenius:2018bai}). {\bf Right:} Gluonic nuclear effects for argon and lead nuclei from EPPS16 (upper panel) and nCTEQ15 (lower panel) at $Q^2=10 \, {\rm GeV}^2$.}
\label{fig:UPCdijet}
\end{figure}

{\bf Ultraperipheral collisions:} The ultraperipheral collisions (UPC) in p-p/p-Pb/Pb-Pb interactions constitute a new avenue of LHC measurements. An interesting observable is the exclusive J/$\psi$ production which at zero-momentum transfer can be related to the square of the nuclear gluon density \cite{Ryskin:1992ui}. The data (e.g. \cite{Khachatryan:2016qhq,LHCb:2018ofh}) also agree nicely with predictions based on such expectations. However, since the momentum transfer is not exactly zero in the measurements, the relation to the gluon comes with an uncertainty and for the moment none of the nuclear-PDF fitting groups have dared to consider this observable in their fits. A theoretically more robust UPC observable from the nuclear-PDF view point is the inclusive dijet photoproduction, measured by ATLAS \cite{ATLAS:2017kwa}. In principle, the yields should be high enough to get quantitative constraints for the nuclear PDFs. This is illustrated in Figure~\ref{fig:UPCdijet} where a prediction from \textsc{pythia} 8 \cite{Helenius:2018bai} using EPPS16 PDFs and the ATLAS kinematics is shown. The EPPS16 errorbands are clearly wider than the expected statistical precision with 1nb$^{-1}$ integrated luminosity and one should be able to resolve shadowing, antishadowing and EMC effects. There are several other inclusive UPC processes which could be measured \cite{Baltz:2007kq} and are sensitive the PDFs, including heavy-flavour and isolated photons.

\vspace{-0.50cm}
\section{On the case for lighter-than-lead ions}

\vspace{-0.30cm}
The Run-II LHC p-Pb data --- with almost 10 times higher statistics than in Run I --- have gradually started to become available. Thus, the global fits of nuclear PDFs will soon be dominated by the measurements involving the lead nucleus (EPPS16 is already). The data on lighter ions come almost exclusively from fixed-target experiments and there is not much variety processwise. In this situation, p-$A$ data with $A \ll 208$ would probably have much more significant impact than, say, running more p-Pb. With lighter ions one can also reach higher luminosities \cite{Jowett} and thus measure rarer processes, or reach the same precision as we have now for p-Pb but in less time. In addition, from the astrophysics view point p-$A$ data with a lighter ion would be useful as e.g. in air the cosmic rays and neutrinos predominantly scatter off from oxygen and nitrogen nuclei. To illustrate a bit how badly the $A$ dependence of PDFs is known, I note that e.g. in the EPPS16 analysis there were only 3 free fit parameters altogether to control the $A$ systematics. Consequently, the bias is presumably significant. The right-hand panels of Figure~\ref{fig:UPCdijet} compare the gluon nuclear modifications for argon and lead as predicted by EPPS16 and nCTEQ15. While in EPPS16 both nuclei look similar, in nCTEQ15 the gluons are even qualitatively very different for these two nuclei. From the nuclear-PDF viewpoint, to constrain the $A$ dependence, it should not matter so much whether it is data on oxygen, argon or some other nucleus, as long as it is clearly smaller than Pb.

Actually, the large-$x$ regime is something one may be able to study also within the fixed-target program of LHCb \cite{Anderlini:2017mom}. For example, there are some D-meson data becoming available \cite{Aaij:2018ogq}, which may shed light on the $A$ dependence of the gluon at large $x$, though there may be other theoretical issues like a possible intrinsic charm component, for instance. Also, the small $\sqrt{s}$ in these measurements may pose theoretical difficulties.

\vspace{-0.30cm}
\section{Summary \& Outlook}

\vspace{-0.20cm}
I have shortly described the present status of the global extractions of nuclear PDFs, and discussed some near-future prospects. Thanks to the increasing variety in data input, the earlier restrictions of imposing flavour-independent nuclear effects for the light quarks have been overcome. Also, as the LHC data are becoming increasingly precise, there are now more pressing motivations for the already ongoing efforts to improve the analyses to an NNLO level. The new LHC Run-II data will bring significant new constraints into the game and in many cases the systematic uncertainties can be expected to dominate over the statistical ones. In this situation the availability of the experimental correlations from one bin to another would be quite important. When planning the future LHC runs, to better understand the $A$ dependence, p-$A$ collisions with some other nucleus than lead would have a significant impact and would be also of great relevance for astrophysics.

\vspace{-0.40cm}
\section*{Acknowledgments}

\vspace{-0.20cm}
\noindent I acknowledge the Academy of Finland projects 297058 and 308301 for support.


\vspace{-0.30cm}
\begin{thebibliography}{99}

\vspace{-0.20cm}
\bibitem{Kovarik:2015cma}
  K.~Kovarik {\it et al.},
  Phys.\ Rev.\ D {\bf 93} (2016) 085037.

\vspace{-0.10cm}
\bibitem{Eskola:2016oht}
  K.~J.~Eskola, P.~Paakkinen, H.~Paukkunen and C.~A.~Salgado,
  Eur.\ Phys.\ J.\ C {\bf 77} (2017) 163.

  \vspace{-0.10cm}
\bibitem{Eskola:2009uj}
  K.~J.~Eskola, H.~Paukkunen and C.~A.~Salgado,
  JHEP {\bf 0904} (2009) 065.
  
  \vspace{-0.10cm}
\bibitem{deFlorian:2011fp}
  D.~de Florian, R.~Sassot, P.~Zurita and M.~Stratmann,
  Phys.\ Rev.\ D {\bf 85} (2012) 074028.
  
  \vspace{-0.10cm}
\bibitem{Khanpour:2016pph}
  H.~Khanpour and S.~Atashbar Tehrani,
  Phys.\ Rev.\ D {\bf 93} (2016) 014026.
  
  \vspace{-0.10cm}
\bibitem{Aschenauer:2017oxs}
  E.~C.~Aschenauer {\it et al.},
  Phys.\ Rev.\ D {\bf 96} (2017) 114005.
  
  \vspace{-0.10cm}
\bibitem{Paukkunen:2017phq}
  H.~Paukkunen,
  PoS DIS {\bf 2017} (2018) 109.

  \vspace{-0.10cm}

\bibitem{Khalek:2018bbv}
  R.~A.~Khalek, J.~J.~Ethier and J.~Rojo,
  arXiv:1811.05858 [hep-ph].

\vspace{-0.10cm}
\bibitem{Sirunyan:2018qel}
  A.~M.~Sirunyan {\it et al.} [CMS Collaboration],
  Phys.\ Rev.\ Lett.\  {\bf 121} (2018) 062002.

  \vspace{-0.10cm}
\bibitem{ATLAS:2018imx}
  The ATLAS collaboration,
  ATLAS-CONF-2018-050.

\vspace{-0.10cm}
\bibitem{jetpaper}
Eskola, Paakkinen, Paukkunen, in preparation.
  
  \vspace{-0.10cm}
\bibitem{Paukkunen:2010qg}
  H.~Paukkunen and C.~A.~Salgado,
  JHEP {\bf 1103} (2011) 071.
  
  \vspace{-0.10cm}
\bibitem{Kusina:2016fxy}
  A.~Kusina {\it et al.},
  Eur.\ Phys.\ J.\ C {\bf 77} (2017) 488.
  
  \vspace{-0.10cm}
\bibitem{Ru:2016wfx}
  P.~Ru, S.~A.~Kulagin, R.~Petti and B.~W.~Zhang,
  Phys.\ Rev.\ D {\bf 94} (2016) 113013.

   \vspace{-0.10cm}
\bibitem{CMS:2018ilq}
  CMS Collaboration,
  CMS-PAS-HIN-17-007.
  
  
  \vspace{-0.10cm}
\bibitem{Guzey:2012jp}
  V.~Guzey, M.~Guzzi, P.~M.~Nadolsky, M.~Strikman and B.~Wang,
  Eur.\ Phys.\ J.\ A {\bf 49} (2013) 35. 
  
  \vspace{-0.10cm}
\bibitem{Brandt:2014vva}
  M.~Brandt, M.~Klasen and F.~König,
  Nucl.\ Phys.\ A {\bf 927} (2014) 78.

    \vspace{-0.10cm}
\bibitem{Helenius:2014qla}
  I.~Helenius, K.~J.~Eskola and H.~Paukkunen,
  JHEP {\bf 1409} (2014) 138.
 
  
  \vspace{-0.10cm}
\bibitem{ATLAS:2017ojy}
  The ATLAS collaboration,
  ATLAS-CONF-2017-072.
  
  \vspace{-0.10cm}
\bibitem{Acharya:2017hdv}
  S.~Acharya {\it et al.} [ALICE Collaboration],
  Phys.\ Lett.\ B {\bf 770} (2017) 459.
  
  \vspace{-0.10cm}
\bibitem{ATLAS:2017dgr}
  The ATLAS collaboration,
  ATLAS-CONF-2017-073.
 
 \vspace{-0.10cm}
\bibitem{Aaij:2017gcy}
  R.~Aaij {\it et al.} [LHCb Collaboration],
  JHEP {\bf 1710} (2017) 090.
  
  \vspace{-0.10cm}
\bibitem{Cacciari:1998it}
  M.~Cacciari, M.~Greco and P.~Nason,
  JHEP {\bf 9805} (1998) 007.

  \vspace{-0.10cm}
\bibitem{Mangano:1991jk}
  M.~L.~Mangano, P.~Nason and G.~Ridolfi,
  Nucl.\ Phys.\ B {\bf 373} (1992) 295.

  \vspace{-0.10cm}
\bibitem{Kniehl:2004fy}
  B.~A.~Kniehl, G.~Kramer, I.~Schienbein and H.~Spiesberger,
  Phys.\ Rev.\ D {\bf 71} (2005) 014018.
  
  \vspace{-0.10cm}
\bibitem{Helenius:2018uul}
  I.~Helenius and H.~Paukkunen,
  JHEP {\bf 1805} (2018) 196.

  \vspace{-0.10cm}
\bibitem{Frixione:2007nw}
  S.~Frixione, P.~Nason and G.~Ridolfi,
  JHEP {\bf 0709} (2007) 126.

  \vspace{-0.10cm}
\bibitem{Aaij:2015bpa}
  R.~Aaij {\it et al.} [LHCb Collaboration],
  JHEP {\bf 1603} (2016) 159.
  
  \vspace{-0.10cm}
\bibitem{Lansberg:2016deg}
  J.~P.~Lansberg and H.~S.~Shao,
  Eur.\ Phys.\ J.\ C {\bf 77} (2017) 1.
  
  \vspace{-0.10cm}
\bibitem{Kusina:2017gkz}
  A.~Kusina, J.~P.~Lansberg, I.~Schienbein and H.~S.~Shao,
  Phys.\ Rev.\ Lett.\  {\bf 121} (2018) 052004.

  \vspace{-0.10cm}
\bibitem{DPaper}
Eskola, Helenius, Paakkinen, Paukkunen, in preparation.
  
  \vspace{-0.10cm}
\bibitem{dEnterria:2013sgr}
  D.~d'Enterria, K.~J.~Eskola, I.~Helenius and H.~Paukkunen,
  Nucl.\ Phys.\ B {\bf 883} (2014) 615.
  
  \vspace{-0.10cm}
\bibitem{Acharya:2018hzf}
  S.~Acharya {\it et al.} [ALICE Collaboration],
  Eur.\ Phys.\ J.\ C {\bf 78} (2018) 624.
  
  \vspace{-0.10cm}
\bibitem{Khachatryan:2015xaa}
  V.~Khachatryan {\it et al.} [CMS Collaboration],
  Eur.\ Phys.\ J.\ C {\bf 75} (2015) 237.

  \vspace{-0.10cm}
\bibitem{Sirunyan:2017xku}
  A.~M.~Sirunyan {\it et al.} [CMS Collaboration],
  Phys.\ Rev.\ Lett.\  {\bf 119} (2017) 242001.
  
  \vspace{-0.10cm}
\bibitem{dEnterria:2015mgr}
  D.~d'Enterria, K.~Krajczár and H.~Paukkunen,
  Phys.\ Lett.\ B {\bf 746} (2015) 64.
  
  \vspace{-0.10cm}
\bibitem{Ryskin:1992ui}
  M.~G.~Ryskin,
  Z.\ Phys.\ C {\bf 57} (1993) 89.
  
  \vspace{-0.10cm}
\bibitem{Khachatryan:2016qhq}
  V.~Khachatryan {\it et al.} [CMS Collaboration],
  Phys.\ Lett.\ B {\bf 772} (2017) 489.
  
  \vspace{-0.10cm}
\bibitem{LHCb:2018ofh}
  LHCb Collaboration,
  LHCb-CONF-2018-003.
  
  \vspace{-0.10cm}
\bibitem{ATLAS:2017kwa}
  The ATLAS collaboration,
  ATLAS-CONF-2017-011.
  
  \vspace{-0.10cm}
\bibitem{Helenius:2018bai}
  I.~Helenius,
  arXiv:1806.07326 [hep-ph].
  
\vspace{-0.10cm}

\bibitem{Baltz:2007kq}
  A.~J.~Baltz {\it et al.},
  Phys.\ Rept.\  {\bf 458} (2008) 1.

\vspace{-0.10cm}  
\bibitem{Jowett}
J.~Jowett, Quark Matter 2018.

\vspace{-0.10cm}

\bibitem{Anderlini:2017mom}
  L.~Anderlini [LHCb Collaboration],
  PoS EPS {\bf -HEP2017} (2017) 152.

  
\vspace{-0.10cm}
\bibitem{Aaij:2018ogq}
  R.~Aaij {\it et al.} [LHCb Collaboration],
  arXiv:1810.07907 [hep-ex].
  
\end{thebibliography}
\end{document}